\documentclass[A4,titlepage,11pt]{article}

\pdfoutput=1

\usepackage{amssymb,amsmath,amsfonts}
\usepackage{epsfig}
\usepackage{ccaption}
\usepackage{graphicx}

\usepackage[
      colorlinks=true,
      linkcolor=blue,
      urlcolor=blue,
      filecolor=black,
      citecolor=red,
      pdfstartview=FitV,
      pdftitle={},
        pdfauthor={Roberto Emparan, Ryotaku Suzuki, Kentaro Tanabe},
        pdfsubject={},
        pdfkeywords={},
        pdfpagemode=None,
        bookmarksopen=true
      ]{hyperref}

\def\clock{{\count0=\time
           \divide\count0 60
           \ifnum\count0<10 0\fi\the\count0
           \multiply\count0 -60 \advance\count0 \time
           :\ifnum\count0<10 0\fi \the\count0
         }}
\newcommand{\timestamp}{{\small\vbox{\hbox{\tt\jobname.tex}
\hbox{\the\day/\the\month/\the\year, \clock}}}}

\newcommand{\lp}{\left(}
\newcommand{\rp}{\right)}

\newcommand{\mc}[1]{\mathcal{#1}}

\newcommand{\beq}{\begin{equation}}
\newcommand{\eeq}{\end{equation}}
\newcommand{\bea}{\begin{eqnarray}}
\newcommand{\eea}{\end{eqnarray}}
\newcommand{\beqa}{\begin{eqnarray}}
\newcommand{\eeqa}{\end{eqnarray}}

\newcommand{\sR}{\mathsf{R}}
\def \a {\alpha}
\numberwithin{equation}{section}
\begin{document}

\begin{titlepage}
%\timestamp
%\rightline{YITP-14-45}
%\rightline{KEK-TH-1741}
%\leftline{}{\timestamp}
\vskip 1cm
\centerline{\LARGE \bf Quasinormal modes of Gauss-Bonnet black holes}
\medskip
\centerline{\LARGE \bf at large $D$}
%\centerline{\LARGE \bf Black hole quasinormal modes at large $D$:}\medskip
%\centerline{\LARGE \bf Decoupling and non-decoupling spectra}
%\vskip 0.15cm
%\centerline{\Large \bf }
\vskip 1.2cm
\centerline{\bf Bin Chen$^{a,b,c}$\footnote{bchen01@pku.edu.cn}, Zhong-Ying Fan$^{c}$\footnote{fanzhy@pku.edu.cn}, Pengcheng Li$^{a}$\footnote{wlpch@pku.edu.cn}, Weicheng Ye$^{a}$\footnote{victorye@pku.edu.cn}}
\vskip 0.5cm
\centerline{\sl $^{a}$Department of Physics and State Key Laboratory of Nuclear Physics and Technology,}
\centerline{\sl Peking University, No.5 Yiheyuan Rd, Beijing 100871, P.R. China}
\smallskip
\centerline{\sl $^{b}$Collaborative Innovation Center of Quantum Matter, No. 5 Yiheyuan Rd,}
\centerline{\sl  Beijing 100871, P. R. China}
\smallskip
\centerline{\sl $^{c}$Center for High Energy Physics, Peking University, No.5 Yiheyuan Rd,}
\centerline{\sl  Beijing 100871, P. R. China}

%\vskip 0.5cm
%\centerline{\small\tt emparan@ub.edu,\, ryotaku@sci.osaka-cu.ac.jp,\, ktanabe@post.kek.jp}

\vskip 1.2cm
\centerline{\bf Abstract} \vskip 0.2cm
\noindent
Einstein's General Relativity theory simplifies dramatically in the limit that the spacetime dimension $D$ is very large. This could still be true in the gravity theory with higher derivative terms. In this paper, as the first step to study the gravity with a  Gauss-Bonnet(GB) term, we compute the quasi-normal modes of the spherically symmetric GB black hole in the large $D$ limit. When the GB parameter is small, we find that the non-decoupling modes are the same as the Schwarzschild case and the decoupled modes are slightly modified by the GB term. However, when the GB parameter is large, we find some novel features. We notice that there are another set of non-decoupling modes due to the appearance of a new plateau in the effective radial potential. Moreover, the effective radial potential for the decoupled vector-type and scalar-type modes becomes more complicated. Nevertheless we manage to compute the frequencies of the these decoupled modes analytically. When the GB parameter is neither very large nor very small, though analytic computation is not possible, the problem is much simplified in the large $D$ expansion and could be numerically treated. We study numerically the vector-type quasinormal modes in this case.

%e study the extension of Einstein's theory to include the Gauss-Bonnet term and derive its implications on the geometry and dynamics of black hole configurations in this article. According to different behaviors of the coupling factor $\alpha$ of the Gauss-Bonnet term, we discover different scenarios of the ``maximally symmetric" black holes in different situations. On one hand, when the Gauss-Bonnet term is dominant or in the subleading order in the Lagrangian, the geometry and dynamics of such ``pure" Gauss-Bonnet black holes is similar to that of Einstein black holes and specifically, the decoupled quasinormal modes of gravitational perturbations in this situation, which reflects true near-horizon geometry, are identical to Einstein's theory at least at the leading order, when the exact value of $\alpha$ does not come into play. On the other hand, when the strength of the Gauss-Bonnet term is comparable with the Einstein term, namely the Lagrangian is a ``hybrid" Lagrangian, the gravitational potential demonstrates a very complicated behavior.

\end{titlepage}
\pagestyle{empty}
\small
%\tableofcontents
\normalsize
\newpage
\pagestyle{plain}
\setcounter{page}{1}

\section{Introduction}

In Einstein's General Relativity, the vacuum equation takes a very simple form
\beq R_{\mu\nu}=0.\eeq
However mathematically concise and beautiful it looks, the equation is a set of coupled highly non-linear partial differential equations. The nonlinearity makes it extremely difficult to analyze. In recent years, Emparan, Suzuki and Tanabe (EST) \cite{kw3,kw4,kw5} proposed an ingenious method called ``Large $D$ Expansion" to study the dynamics of the black holes. EST considers the limit that the spacetime dimension is very large and develops a systematic way to do $1/D$ expansion. This method was inspired by the large $N$
expansion of $SU(N)$ gauge theories\cite{kw1,kw2}. Extended objects called strings are formulated in the large $N$ expansion of Yang-Mills theories and the counterpart of the string in the Large $D$ expansion of gravity is the black hole. A Schwarzschild black hole of a Schwarzschild radius $r_0$ in $D$ spacetime dimensions is described by the metric \cite{kw20}
\beq ds^2=-\Big(1-\big(\frac{r_0}{r}\big)^{D-3}\Big)dt^2+\frac{dr^2}{\Big(1-\big(\frac{r_0}{r}\big)^{D-3}\Big)}+r^2 d\Omega_{D-2}^2. \label{metric} \eeq
We can see that the geometry of a black hole in $D$ spacetime dimensions is non-trivial only in a distance $\frac{r_0}{D-3}$ away from its event horizon outside of which the geometry can be essentially taken as the Minkowskian spacetime. Therefore, the black holes can be regarded as non-interacting ``particles" of finite radius but vanishingly small cross sections \cite{kw3}. Thus, the focus of EST's work has been mainly on these non-perturbative extended objects, while the black branes and the membranes have also been considered in their work and the following works by other groups\footnote{For another large D limit, see \cite{Giribet:2013wia}.} \cite{kw3,kw4,kw5,kw6,kw7,kw8,kw9,kw10}.

Most importantly, EST have developed a systematic method of computing the quasinormal modes by the $1/D$ expansion and obtained the results in perfect agreement with previous numerical results \cite{kw6,kw8,kw9}. They found that there are two kinds of quasinormal modes, the non-decoupling ones and the decoupled ones. The non-decoupling ones are non-renormalizable in the near horizon geometry, and such modes  have frequencies of order $\frac{D}{r_0}$.  These modes, however, are universally shared among  all spherically static black holes since they essentially reflect the asymptotic flatness of the black hole so that they carry little information about the black hole geometry. Besides, there are decoupled modes localized within the near horizon region, with their frequencies being of order $\frac{1}{r_0}$.  In contrast to the non-decoupling modes, the decoupled modes is tightly related to the specific black hole geometry beyond the leading large $D$ limit. Therefore, their values at the higher orders exhibit detailed near-horizon properties of a specific black hole.

The quantum corrections to the classical general relativity implies the existence of the higher curvature terms. Among the higher curvature terms, the so-called Gauss-Bonnet(GB) term is of particular interest. It is made up of the quadratic terms in curvature, and appears as the leading correction in string theory\cite{Zwiebach:1985uq,kw11}.  This term is a simple topological term when $D = 4$, and becomes physically relevant only when $D\geq5$. Including the Gauss-Bonnet term into the gravity action, we get
\beq
I=\frac{1}{16\pi G_D}\int d^D\sqrt{-g}\biggl(R+\alpha(R_{abcd}R^{abcd}-4R_{cd}R^{cd}+R^2)\biggl),
\eeq
where $\a$ is a parameter for the GB term. This action describes the so-called Einstein-Gauss-Bonnet gravity or simply Gauss-Bonnet gravity. One nice thing about this action is that the equation of motion is still of second order and there is no ghost. Another nice thing is that there are well-known black hole solutions in this theory.

%No-ghost condition and other physical constraints can only determine the basic form of the Gauss-Bonnet term, leaving the coupling factor $\alpha$ of this term unconstrained, except an estimation based on the size of strings predicted by the string theory. Therefore, the contribution of the Gauss-Bonnet term or, one step further, even higher order corrections is basically unknown to us, even though we may simply assume its contribution to be small by naive physical intuitions. It is worth noting that probing our own universe will not give us information about the strength of this term, not until all the way down to an extremely high energy scale when extra dimensions or something beyond Standard Model come into play.

A natural generalization of EST's work is to perform a large $D$ expansion in the  Gauss-Bonnet gravity theory. As the first step, we would like to compute the quasi-normal modes of a spherically symmetric GB black hole in the large $D$ expansions. The master equations for the scalar-, vector- and tensor-type perturbations  have been computed in \cite{kw17,kw18, kw15, kw19}.  A complete numerical analysis of the evolution of the gravitational perturbations for $D$-dimensional Gauss-Bonnet black holes with $D = 5\sim 11$ was performed by Konoplya \cite{kw19}, and the stability and instability regions have been determined comprehensively there.
The aim of our work is to perform a systematic calculation of quasinormal modes in the large $D$ expansion. %The method we employ is straightforward.

The Gauss-Bonnet term could be originated from the string theory which might restrict the value of the parameter $\a$. However, in this work, we just focus on the Gauss-Bonnet gravity, without restricting the value of $\a$. In the large $D$ expansion, the value of the parameter $\a$ determines the contribution of the GB term.  It is easy to see that  the Riemann tensor $R_{\mu\nu}$ scales as $D^2$, and therefore the Einstein term also scales as $D^2$ while the Gauss-Bonnet term scales as $D^4$! It is natural to  assume that the value of $\alpha$ does not change with $D$, leading to a theory that is dominant by the Gauss-Bonnet term at large $D$. Under such circumstance the magnitude of the Einstein term is of two less orders  than that of the Gauss-Bonnet term, and we can regard the theory as a ``pure" Gauss-Bonnet theory at large $D$ which includes only the Gauss-Bonnet term, plus a small perturbation.  The situation when $\alpha \sim O(D^{-1})$ or larger is similar and we can just take $\alpha\sim O(D^0)$ as an illustrative example. On the other hand, when $\alpha$ scales as $D^{-3}$ or less, the Einstein term dominates, and the black hole can be regarded as an Einstein black hole with small perturbations from the Gauss-Bonnet term. Analytical results can be obtained for both the small and large $\a$ cases. The situation becomes complicated if
 $\alpha$ scales as $D^{-2}$. In this case the magnitude of the Einstein term is the same  as that of the Gauss-Bonnet term. Although in the large $D$ expansion  the problem can still be simplified dramatically, the analytical treatment is not feasible and the numerical calculations have to be carried out. %We study decoupled quasinormal modes of vector perturbations in this case half analytically half numerically, and obtain results sensitive to the exact value of $\tilde{\alpha}$, which demonstrates a ``competition" of the two terms. When $\tilde{\alpha}$ is around $0.65(r_0^2)$, quasinormal modes display the largest deviation from quasinormal modes of both ``pure" Gauss-Bonnet black holes and Einstein black holes.

The paper is organized as follows. In Sec.~\ref{section2} we introduce  the geometry of the Einstein-Gauss-Bonnet black hole in the large $D$ expansion. In Sec.~\ref{section3} we discuss briefly the quasinormal modes for a minimally-coupled scalar field. In  Sec.~\ref{section4} and Sec.~\ref{section5} we study the non-decoupling and decoupled quasinormal modes respectively. In Appendix,  we give numerical results for the decoupled vector-type quasinormal modes of ``hybrid" Gauss-Bonnet black holes at the leading order. %in the Appendix.

\section{Basic geometry}\label{section2}

The metric of a spherically symmetric and static black hole in the Einstein-Gauss-Bonnet gravity could be  written as\cite{kw11}
\begin{equation}
ds^2=-f(r)dt^2+\frac{dr^2}{f(r)}+r^2d\Omega_{n+2}^2, \label{GBBH}
\end{equation}
\begin{equation}
f(r)=1+\frac{r^2}{\alpha(D-3)(D-4)}(1-q(r)),\qquad q(r)=\sqrt{1+\frac{4\alpha(D-3)(D-4)\mu}{(D-2)r^{D-1}}},
\end{equation}
where $\mu$ is the mass of the black hole. The horizon is at $r=r_H$ which is related to the mass $\mu$ by the relation
\beq
\mu=\frac{(D-2)r_H^{D-3}}{4}\biggl(2+\frac{\alpha(D-3)(D-4)}{r_H^2}\biggl).
\eeq
For convenience we can set $r_H=1$ and introduce an useful quantity \beq
\tilde{\alpha}\equiv\alpha (D-3)(D-4)/2.\eeq
 In terms of $\tilde{\alpha}$ the $f(r)$ and $q(r)$ can be expressed as
\beq
f(r)=1+\frac{r^2}{2\tilde{\alpha}}(1-q(r)),\qquad q(r)=\sqrt{1+\frac{4\tilde{\alpha}(1+\tilde{\alpha})}{r^{D-1}}}.
\eeq
In order to discuss the large $D$ expansion, we introduce an expansion parameters
\begin{equation}
\qquad n\equiv D-3,
\end{equation}
and let
\beq
\sR\equiv\big(\frac{r}{r_H}\big)^n.\eeq

\subsection{Small $\tilde{\alpha}$}

When $\tilde{\alpha}$ is small, for example $\tilde \alpha \sim \mc O(1/n)$, the second term in $q(r)^2$ is very small so $q(r)$ can be expanded as a power-series of $\tilde{\alpha}$. This is always possible because we are interested in the geometry outside the horizon such that $r>r_H=1$ and therefore $4\tilde{\alpha}(1+\tilde{\alpha})\ll1$. In order to precisely represent the ``smallness" of  $\tilde{\alpha}$, we introduce a new parameter $\beta\equiv n\tilde{\alpha}$, which is of order one  $\beta\sim\mc O(1)$. Using $1/n$ expansion the above formulas can be expanded as

\begin{equation}
q(r)=1+\frac{1}{n}\frac{2\beta}{\sR}+\frac{1}{n^2}\frac{2\beta(\beta(\sR-1)-2\sR\ln\sR)}{\sR^2}+\mc O{(\frac{1}{n^3})},
\end{equation}
\beq\label{eq18}
f(r)=1-\frac{1}{\sR}+\frac{1}{n}\frac{\beta(1-\sR)}{\sR^2}+\frac{1}{n^2}\frac{2\beta(\beta(\sR-1)-\sR\ln\sR)}{\sR^3}+\mc O{(\frac{1}{n^3})},
\eeq
In this case the near region is $r-r_H\ll r_H$, or $\sR\ll e^n $ and the far region is $\sR\gg 1$. Obviously, when $\beta\to 0$ we recover the Schwarzschild case in pure Einstein gravity.

We could expect that in the case that $\tilde \a$ is small, the corrections originating from the Gauss-Bonnet term must be small. From the expansion in the function $f(r)$, the effects from the Gauss-Bonnet term  should only be reflected at the $1/n$ order or even higher order terms.  In the limit $\beta\to 0$ we should reproduce the results in the Schwartzschild black hole in the Einstein gravity.

\subsection{Large $\tilde{\alpha}$}

When $\tilde{\alpha}$ is large enough, for example $\tilde \alpha \sim \mc O(n^2)$ and $\alpha\sim\mc O(1)$, in the region where $\sR\ll n^4$, the second term in $q(r)^2$ dominates and we could expand it in series of $1/n$, so the forms of $q(r)$ and $f(r)$ are expanded as
\beq
q(r)=n^2\frac{\alpha}{\sqrt{\sR}}-n \frac{\alpha}{\sqrt{\sR}}(1+\ln\sR)+ \frac{1}{2\sqrt{\sR}}(2+2\alpha+\alpha\ln\sR+\alpha(\ln R)^2)+\mc O(\frac{1}{n}),
\eeq
\beq\label{eq19}
f(r)=1-\frac{1}{\sqrt{\sR}}-\frac1 n\frac{\ln\sR}{\sqrt{\sR}}+\frac{1}{ n^2} \frac{-2+2\sqrt{\sR}-\alpha(\ln\sR)^2}{2\sqrt{\sR}\alpha}+ \mc O(\frac{1}{n^3}).
\eeq
The validity of the expansion requires that $\sR\ll n^4$, which we will refer to as the near region although  it is smaller than the usual near region where $r-r_H\ll r_H$. This is all right since it still has overlap with the far region $\sR\gg1$ .

When $\tilde{\alpha}$ is a little smaller, e.g.$\sim \mc O(n)$ or even larger e.g.$\sim \mc O(n^3)$, the discussion on the $1/n$ expansion is similar. The only difference is that the near region becomes smaller $\sR\ll n^2$ or larger $\sR\ll n^6$. Therefore we will treat  $\tilde{\alpha}\sim \mc O(n^2)$ as a typical example for the large $\tilde \alpha$ case and discuss it explicitly.

However, when $\tilde{\alpha}$ takes an intermediate value, i.e. $\tilde{\alpha} \sim \mc O(1)$, even in the large $D$ expansion the metric is too complicated for us to compute the  quasinormal modes analytically. In this case, the leading order form of $f(r)$ is given by
\beq
f(r)=1+\frac{1}{2\tilde{\alpha}}-\frac{1}{2\tilde{\alpha}}\sqrt{1+\frac{4\tilde{\alpha}(1+\tilde{\alpha})}{\sR}},
\eeq
which is quite different from \eqref{eq18} or \eqref{eq19}. As a result there are different  spectrums for the decoupled modes  which are not universal and depend on the specific black hole geometry. In the appendix we present the numerical result of vector-type quasinormal modes at leading order to show this point.

\section{The quasinormal modes for a scalar field}\label{section3}

As the first step, let us consider a minimally-coupled scalar in the black hole background. As the black hole geometry is spherically symmetric, the scalar wavefunction
could be decomposed into the following form
\beq
\Phi=e^{-i\omega t}\phi(r)Y_{l,m}(\theta,\varphi)
\eeq
where $\omega$ is the frequency and $Y_{l,m}$ is the spherical harmonic function.
The differential equation for the radial function is

\begin{equation}\label{eq31}
\frac{d}{d \sR}\biggl(f(r)\sR^2\frac{d}{d\sR}\phi(\sR)\biggl)+\frac{\hat{\omega}^2 \sR^{2/n}}{f(r)}\phi(\sR)-(\omega_{c}^2-\frac{1}{4})\phi(\sR)=0.
\end{equation}
where $\omega_c=\hat{\ell}+\frac{1}{2}$, $\hat{\ell}=\ell/n$, $\hat{\omega}=\omega/n$. The above equation can be recast into a master equation of the form

\begin{equation}\label{equation32}
\biggl(\frac{d^2}{dr_*^2}+\omega^2-V(r_*)\biggl)\psi(r)=0,
\end{equation}
where
\beq
r_*=\int\frac{dr}{f(r)}, \qquad \psi(r)=\sR^{\frac{1}{2}+\frac{1}{2n}}\phi(r),
\eeq

\begin{equation}
V(r_*)=\frac{f(r)}{4r^2}((n^2-1)f(r)+4l(n+l)+2(n+1)rf'(r)).
\end{equation}
The height of the potential is $V^{\text{max}}=n^2 \omega_c^2$. Since the potential varies slowly in the overlapping zone,  we can treat it as a constant. As a consequence,  the differential equation \eqref{equation32} in the overlapping region takes the form
\beq
\biggl(\frac{d^2}{d(\ln\sR)^2}+\hat{\omega}^2-\omega_c^2 \biggl)\psi=0.
\eeq
In fact this form of the radial equation is independent of the value of $\tilde{\alpha}$, although the range of the overlapping region depends on $\tilde \alpha$. If $\hat \omega \neq \omega_c$, the solution of this equation is
\beq\label{equationbc3}
\psi=A_+\,\sR^{\sqrt{\omega_c^2-\hat{\omega}^2}}+A_-\,\sR^{-\sqrt{\omega_c^2-\hat{\omega}^2}},
\eeq
while if  $\hat{\omega}=\omega_c$, the solution is of the form
\beq\label{equationbc1}
\psi=A+B\ln \sR,
\eeq
where $A_\pm$, $A,B$ are integration constants.

The  quasinormal modes are the solutions of \eqref{equation32}, which  should satisfy the ingoing boundary condition at the event horizon and the outgoing boundary condition at the infinity. At the horizon this requires
\beq
\psi(\sR)=(\sR-1)^{-i \hat{\omega}}\phi_s(\sR),
\eeq
when $\tilde{\alpha}$ is small and
\beq\label{equationbc2}
\psi(\sR)=(\sqrt{\sR}-1)^{-2i \hat{\omega}}\phi_s(\sR),
\eeq
when $\tilde{\alpha}$ is large. Here $\phi_s(\sR)$ is some regular function at $\sR=1$.

The strategy to find the quasinormal modes is to solve the differential equation of the perturbation in the far region and the near region with appropriate boundary conditions. The matching of the solutions in the overlapping region then determines the quasinormal modes.

\subsection{Near-region solutions}
\subsubsection{Small $\tilde{\alpha}$}

For a small $\tilde{\alpha}$, consider the leading order in the $1/n$ expansion in the near region, from the behavior of $f(r)$ in \eqref{eq18} we see that the differential equation  \eqref{eq31} is exactly the same as the one in the Schwarzschild black hole, which is of a hypergeometric type. Hence the solution that satisfies the ingoing boundary condition at the horizon is exactly the same as the result in \cite{kw3},
\beq
\psi(\sR)=(\sR-1)^{-i\hat{\omega}}\sqrt{\sR}\,{}_2F_1(a, b, a+b; 1-\sR),
\eeq
where
\begin{equation}
\begin{split}
a=\frac{1}{2}+\sqrt{\omega_c^2-\hat{\omega}^2}-i\hat{\omega},\qquad b=\frac{1}{2}-\sqrt{\omega_c^2-\hat{\omega}^2}-i\hat{\omega}.
\end{split}
\end{equation}

\subsubsection{Large $\tilde{\alpha}$}

For a large $\tilde{\alpha}$, up to the $1/n^0$ order, the radial equation is simplified to be
\begin{equation}
\frac{d}{d \sR}\biggl(\sR^2-\sR^{3/2}\biggl)\frac{d}{d\sR}\phi(\sR)+\frac{\hat{\omega}^2}{1-\sR^{-1/2}}\phi(\sR)-(\omega_{c}^2-\frac{1}{4})\phi(\sR)=0.
\end{equation}
The solution is
\begin{equation}
\begin{split}
\phi(\sR)=&A_1(\sqrt{\sR}-1)^{-2i\hat{\omega}} \,{}_2F_1(a,b,a+b;1-\sqrt{\sR})
\\&+A_2(\sqrt{\sR}-1)^{2i\hat{\omega}}\, {}_2F_1(2-a,2-b,3-a-b;1-\sqrt{\sR}),
\end{split}
\end{equation}
where
\begin{equation}
a=1-2i\hat{\omega}+2\sqrt{w_c^2-\hat{\omega}^2}, \qquad b=1-2i\hat{\omega}-2\sqrt{w_c^2-\hat{\omega}^2}.
\end{equation}
The boundary condition \eqref{equationbc2} selects the solution to be
\beq
\psi(\sR)=(\sqrt{\sR}-1)^{-2i\hat{\omega}} \sqrt{\sR}\,{}_2F_1(a,b,a+b-1;1-\sqrt{\sR}).
\eeq

As discussed in \cite{kw8}, the only information that we need from the solution is their large $\sR$ behavior in the overlapping region where $\sR\gg1$. It is easy to find that for a general $\hat{\omega}$ no matter what value $\tilde{\alpha}$ takes there is always
\beq\label{equation315}
\biggl|\frac{A_+}{A_-}\biggl|=\mc O(1).
\eeq
When $\hat{\omega}=\omega_c$, the solution should be in match with \eqref{equationbc1} in the overlapping region, leading to
\beq\label{equation316}
\biggl|\frac{A}{B}\biggl|=\mc O(1).
\eeq

\subsection{Far-region solutions}
In the far-region, $1/\sR$ is exponentially small. Thus we can set $f=1$ no matter what value $\tilde{\alpha}$ takes and the radial equation \eqref{equation32} is exactly the same as the one in the Minkowski spacetime, so the solution is just the Hankel functions \cite{kw8}
\beq
\psi(r)=\sqrt{r}\,H_{n\omega_c}^{(1)}(\omega\,r).
\eeq
 Following the discussion in \cite{kw8}, in the overlapping region, in terms of the coordinate $\sR$ the solution takes the form \eqref{equationbc3} or \eqref{equationbc1}.

\subsection{Quasinormal modes}

As we have seen, the  far-region solution is exactly the same as the one in the Schwarzschild black holes whatever $\tilde{\alpha}$ takes. On the other hand in the near region, the radial equation for a small $\tilde{\alpha}$  is identical to the scalar equation in the Schwarzschild black hole background. But for a large $\tilde{\alpha}$ the radial equation and its solution in the near region is different. Nevertheless the useful information in the overlapping region is encoded  in Eqs. \eqref{equation315} and \eqref{equation316}, the same as the ones in the Schwarzschild case.

If we try to paste the solutions in all regions satisfying the appropriate boundary conditions, we see that even though the solution in the near region could be different, the solution in the overlapping region is the same as the Einstein gravity. Consequently  we conclude that the quasinormal modes for a scalar field in the Gauss-Bonnet black hole \eqref{GBBH} are completely the same as the ones in the Schwarzschild black hole background.

\section{Non-decoupling modes for gravitational perturbations}\label{section4}

In the last section, we discussed the quasinormal modes of a scalar field in the Gauss-Bonnet black hole background. The scalar field is taken as a probe and is minimally coupled to the gravity. It could only probe the geometry of the background but cannot
see the dynamics of the Gauss-Bonnet gravity. In this section, we discuss the gravitational perturbation in the Gauss-Bonnet gravity. This may allow us to investigate the dynamics of the theory.

  The linearized gravitational fluctuations could be classified according to their transformation properties under the rotation group: scalar-type(S), vector-type(V) and tensor-type(T) gravitational perturbations. Each type of the perturbations satisfies the master equation of the form
\begin{equation}\label{equation31}
\biggl(\frac{d^2}{dr_*^2}+\omega^2-V_s\biggl)\Psi_s=0,
\end{equation}
where $s=S, V, T$ denotes  three types of the perturbations. The potentials in \eqref{equation31} depend on the types of the perturbation \cite{kw19}
 \begin{equation}
V_T(r)=f(r)\frac{\lambda}{r^2}\biggl(3-\frac{B(r)}{A(r)}\biggl)+\frac{1}{\sqrt{r^{D-2}A(r)q(r)}}\frac{d^2}{dr_*^2}\sqrt{r^{D-2}A(r)q(r)},
\end{equation}

\begin{equation}
V_V(r)=f(r)\frac{(D-2)c}{r^2}A(r)+\sqrt{r^{D-2}A(r)q(r)}\frac{d^2}{dr_*^2}\frac{1}{\sqrt{r^{D-2}A(r)q(r)}},
\end{equation}

\begin{equation}
V_S(r)=\frac{f(r)U(r)}{64r^2(D-3)^2A(r)^2q(r)^8(4cq(r)+(D-1)^2(q(r)^2-1)^2)},
\end{equation}

and
\begin{equation}
A(r)=\frac{1}{q(r)^2}\biggl(\frac{1}{2}+\frac{1}{D-3}\biggl)+\biggl(\frac{1}{2}-\frac{1}{D-3}\biggl),
\end{equation}
\begin{equation}
B(r)=A(r)^2\biggl(1+\frac{1}{D-4}\biggl)+\biggl(1-\frac{1}{D-4}\biggl),
\end{equation}
\begin{equation}
H=\frac{r^2}{\tilde{\alpha}}, \qquad \lambda=(D-2)(c+1)=l(l+D-3),
\end{equation}
\begin{equation}
\begin{split}
&U(r)=5 (D - 1)^6 H^2 (1 + H) - 3 (D - 1)^5 H ((D - 1) H^2 + 24 c (1 + H)) q(r)  \\ +
 &2 (D - 1)^4 (24 c (D - 1) H^2 +   168 c^2 (1 + H) - (D - 1) H^2 (-3 + 5 H + 7 D (1 + H))) q(r)^2 \\+
& 2 (D - 1)^4 H (-184 c^2 + (D - 1) (13 + D) H^2 +
   c (-84 + 44 H + 84 D (1 + H))) q(r)^3 \\ +
   &    (D - 1)^3 (384 c^3 - 48 c (2 + D (3 D - 5)) H^2 +
    192 c^2 (-11 + D + (-15 + D) H)  \\
 +& (D - 1) H^2 (-3 (7 + 55 H) +
     D (26 + 106 H + 7 D (1 + H)))) q(r)^4  \\
 +&       (D - 1)^3 H (-64 c^2 (D - 38) + (D - 1) (71 + D (7 D - 90)) H^2  \\ +
 & 16 c (303 + 255 H + 13 D^2 (1 + H) - 2 D (73 + 81 H))) q(r)^5 \\ +
 &4 (D - 1)^2 (96 c^3 (-7 + D) -
  8 c (D - 1) (145 - 74 D + 6 D^2) H^2  \\-
  &  8 c^2 (9 - 175 H + D (-58 - 34 H + 11 D (1 + H))) + (D - 1) H^2 (-5 (79 + 23 H)  \\ +
   &    D (5 (57 + 41 H) + D (-81 - 89 H + 7 D (1 + H))))) q(r)^6 \\-
& 4 (D - 1)^2 H (8 c^2 (43 + (72 - 13 D) D) + (D - 1) (-63 + D (99 + D (-49 + 5 D))) H^2  \\ +
 &   4 c (321 + 465 H + D (121 - 39 H + D (-123 - 107 H + 17 D (1 + H))))) q(r)^7  \\+
  &     (D - 1) (128 c^3 (-9 + D) (D - 5) + 32 c (D - 1) (246 + D (9 + D (-55 + 8 D))) H^2  \\ +
   & 64 c^2 (D - 5) (D^2 - 3 + (49 + (D - 4) D) H)  \\ -
  &(D - 1) H^2 (1173 + 565 H + D (-4 (997 + 349 H) + D (6 (393 + 217 H) + D (-548 - 452 R + 45 D (1 + H)))))) q(r)^8  \\ +
  &  (D - 1) H (-64 c^2 (D - 5) (36 + D (-13 + 3 D)) + (D - 1) (635 + D (-1204 + 3 D (294 + D (-92 + 9 D)))) H^2  \\ -
   & 8 c (D - 5) (63 + 31 H + D (127 + 191 H + D (-47 + D + (-79 + D) H)))) q(r)^9  \\ +
 &2 (D - 5) (64 c^3 (D - 5) (D - 3) + 8 c (D - 1) (-27 + D (141 + (-43 + D) D)) H^2 \\ +
  &  8 c^2 (D - 5) (-3 + 77 H + D (D - 2 + (D - 18) H)) + (D - 1)^2 H^2 (-33 (H - 7)  \\ +
   &    D (59 + 43 H + D (-59 - 35 H + 9 D (1 + H))))) q(r)^{10}  \\ -
& 2 (D - 5) H (24 c^2 (-11 + D) (D - 5) (D - 3) + (D - 1)^2 (-65 + D (81 + D (7 D - 39))) H^2  \\ +
 &   12 c (-7 + D) (D - 5) (D - 3) (D - 1) (1 + H)) q(r)^{11}  \\ +
  &  (D - 5)^2 (-1 + D) H^2 (16 c (26 + (D - 9) D) + (D - 1) (77 - 3 H + D (-18 + D + (D - 2) H))) q(r)^{12}  \\ +
   & (D - 5)^2 (D - 3)^2 (D - 1)^2 H^3 q(r)^{13},
\end{split}
\end{equation}

The discussion on the non-decoupling quasinormal modes is similar to the one for the scalar field in the previous section. In order to find the quasinormal modes, one need to solve the master equation in two different regions and then match  them in the overlapping region.

\subsection{Small $\tilde{\alpha}$ }

 Up to the leading order in $1/n$, the metric is the same as the one in the Schwarzschild case, so we could expect that the quasinormal modes are the same as long as we only keep to leading order. Although the three potential forms are complicated, the GB effect appears only at the next-to-leading order. Actually  the leading order form of the three potentials are
\beq
V_T=\frac{(\sR-1)(4\omega_c^2\sR+1)}{4\sR^2},
\eeq
\beq
V_V=\frac{(\sR-1)(4\omega_c^2\sR-3)}{4\sR^2},
\eeq
\beq
V_S=\frac{(\sR-1)(1+(1+8\hat{\ell}(\hat{\ell}+1))\sR-12\hat{\ell}(\hat{\ell}+1)(\hat{\ell}(\hat{\ell}+1)+1)\sR^2+4\hat{\ell}^2(\hat{\ell}+1)^2(1+2\hat{\ell})^2\sR^3)}{4\sR^2(1+2\hat{\ell}\sR+2\hat{\ell}^2\sR)^2}.
\eeq
All of them are independent of $\beta$  so that  the GB term has no effect on the non-decoupling quasinormal modes in the small $\tilde{\alpha}$ case.

\subsection{Large $\tilde{\alpha}$ }
\begin{figure}[t]\label{fig1}
 \begin{center}
  \includegraphics[width=.5\textwidth,angle=0]{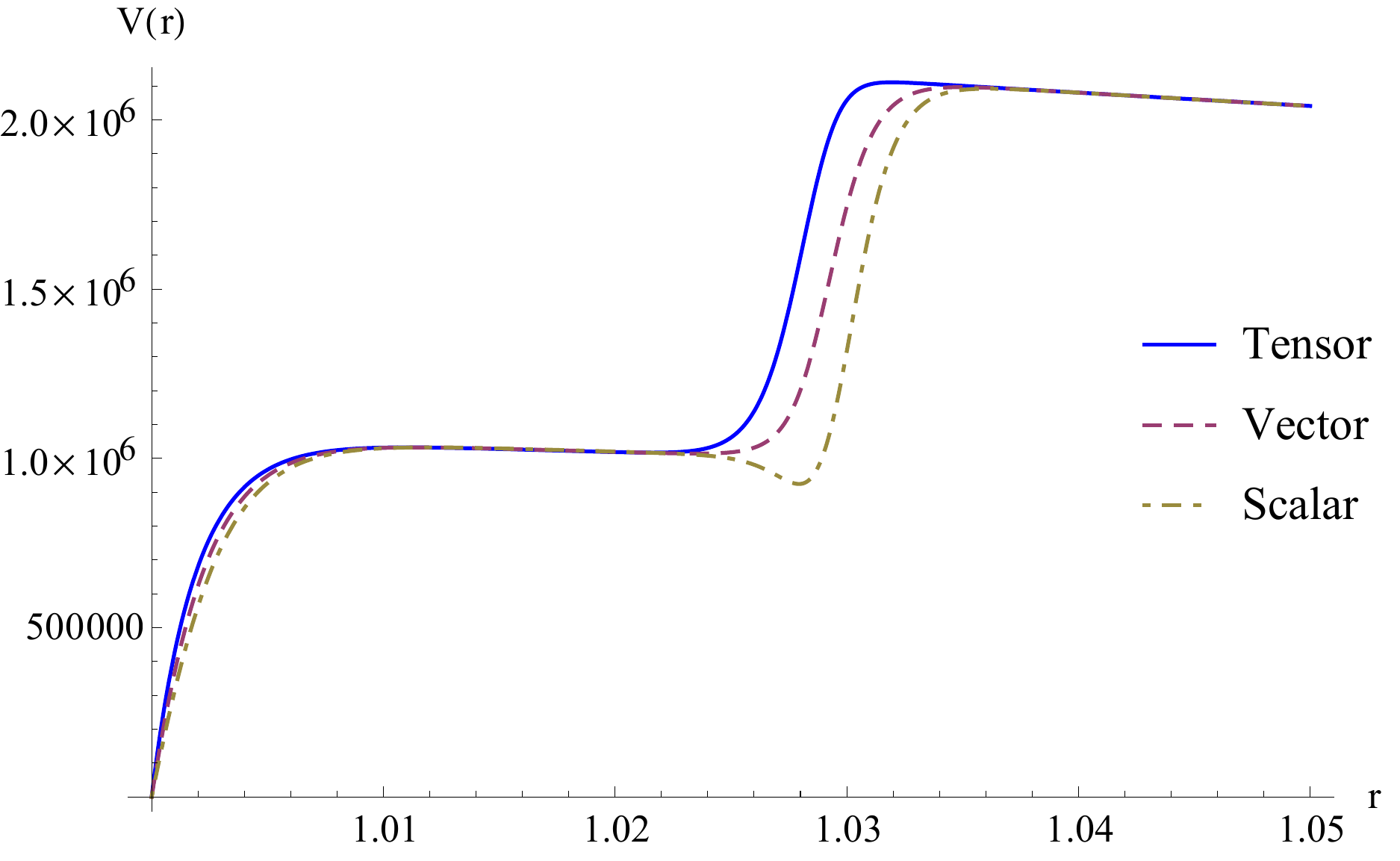}
   \end{center}
 \vspace{-5mm}
 \caption{\small Radial  potential $V_s(r)$ for the perturbations of different types in the Einstein-Gauss-Bonnet black hole for $n=10^3$, $\ell=10^3$ and $\tilde{\alpha}=10^6$. The horizon is at $r=1$. We use the coding solid, dashed, and dot-dashed lines to denote the potentials for the tensor-, vector- and scalar-type gravitational perturbations respectively. }

\end{figure}
In Fig.~\ref{fig1}, we show the potentials  for different types of gravitational perturbation in the very large $n$ limit. Here we choose $n=10^3$, $\tilde{\alpha}=10^6$, and $\ell=10^3$. From Fig.~\ref{fig1}, we find that  there are two plateaux,  a fact that  is very different from the Schwarzschild case. The lower one is unique for the Einstein-Gauss-Bonnet gravity and it disappears in the limit that $\tilde{\alpha}$ goes to zero. Its height  is $(1+8\hat{\ell}+8\hat{\ell}^2)n^2/16$. It is in the range of the near horizon region $\sR\ll\tilde{\alpha}^2$ we defined before  so that it is suitable for the $1/n$ expansion. The higher one is beyond the near horizon region, and its height is  $n^2\omega_c^2$,  the same as the one in the Schwarzschild case, since the Gauss-Bonnet black hole we considered is also asymptotically flat. Recall that the non-decoupling modes are non-normalizable in the near horizon geometry. The real part of the frequency of the non-decoupling quasinormal modes is lower than the maximum of the potential. As now there are two separated plateaux in the potential,  there might be two different sets of  non-decoupling modes. Let us work them out in detail.

\subsubsection{Lower plateau}
Let us focus on the lower plateau first. Up to the leading order in the $1/n$ expansion the heights of the three potentials are the same, so in this case  the basic form of the solutions of the master  equation \eqref{equation31} in the overlapping region must be
\beq\label{equation312}
\Psi_s=A_+\,\sR^{\sqrt{\bar{\omega}_c^2-\hat{\omega}^2}}+A_-\,\sR^{-\sqrt{\bar{\omega}_c^2-\hat{\omega}^2}},
\eeq
where we have defined a new quantity $\bar{\omega}_c=\sqrt{1+8\hat{\ell}+8\hat{\ell}^2}/4$. When $\hat{\omega}=\bar{\omega}_c$ there is
\beq\label{equation313}
\Psi_s=A+B\ln \sR.
\eeq

For different types of the perturbations, the potentials in the master equation take different forms. Up to the leading order they are respectively
\beq
V_T=\frac{(\sqrt{\sR}-1)(16\bar{\omega}_c^2\sqrt{\sR}+1)}{16\sR},
\eeq
\beq
V_V=\frac{(\sqrt{\sR}-1)(16\bar{\omega}_c^2\sqrt{\sR}-3)}{16\sR},
\eeq
\beq
V_S=\frac{(\sqrt{\sR}-1)(1+(1+16\hat{\ell}+16\hat{\ell}^2)\sqrt{\sR}-24\hat{\ell}(1+3\hat{\ell}+4\hat{\ell}^2+2\hat{\ell}^3)\sR
+16\hat{\ell}^2(1+\hat{\ell})^2(1+8\hat{\ell}+8\hat{\ell}^2)\sR^{3/2})}{16\sR(1+4\hat{\ell}\sqrt{\sR}+4\hat{\ell}^2\sqrt{\sR})^2}.
\eeq
For the tensor- and vector-perturbations, their equations are of hypergeometric types. Taking into account the boundary condition at the horizon the solutions are respectively
\beq
\Psi_T(\sR)=(\sqrt{\sR}-1)^{-i2\hat{\omega}}\sR^{1/4}\,{}_2F_1(a,b,a+b;1-\sR),
\eeq
\beq
\Psi_V(\sR)=(\sqrt{\sR}-1)^{-i\hat{\omega}}\sR^{3/4}\,{}_2F_1(a+1,b+1,1+a+b;1-\sR),
\eeq
where
\begin{equation}
\begin{split}
a=\frac{1}{2}+2\sqrt{\bar{\omega}_c^2-\hat{\omega}^2}-i2\hat{\omega},\qquad b=\frac{1}{2}-2\sqrt{\bar{\omega}_c^2-\hat{\omega}^2}-i2\hat{\omega}.
\end{split}
\end{equation}
For the scalar-type potential the equation is more complicated. As in the case discussed in \cite{kw8}, the scalar solution can be expressed as some differential operators acting on a hypergeometric function.

The only information that we need from the solutions is their large $\sR$ behaviors in the overlapping region where $\sR\gg1$. It is easy to find that for general $\hat{\omega}$ , we have
\beq
\biggl|\frac{A_+}{A_-}\biggl|=\mc O(1),
\eeq
while for $\hat{\omega}=\bar{\omega}_c$,
\beq
\biggl|\frac{A}{B}\biggl|=\mc O(1).
\eeq

Similarly in the far-region we may set $f=1$, and the outgoing solution is the Hankel function
\beq\label{equation322}
\Psi_s=\sqrt{r}H_{n\bar{\omega}_c}(\omega r).
\eeq
In terms of the coordinate $\sR$ the solution can take the form \eqref{equation312} or \eqref{equation313}.

%\subsubsection{Non-decoupling quasinormal modes}

The matching of the near- and far-region solutions  give the non-decoupling modes whose frequency are of order $\sim \mc O(n)$. As discussed in \cite{kw8}, the least-damped modes have analytic
expressions and the case of higher overtones could be described numerically. Here, we present the real part and the imaginary part of the least-damped mode frequencies as follows
\beq
\omega_{\text{R}}=\frac{n}{4}\sqrt{1+8\hat{\ell}+8\hat{\ell}^2}-\frac{a_k}{4}n^{1/3}(1+8\hat{\ell}+8\hat{\ell}^2)^{1/6},
\eeq
and
\beq
\omega_{\text{I}}=-\frac{\sqrt{3}a_k}{4}n^{1/3}(1+8\hat{\ell}+8\hat{\ell}^2)^{1/6},
\eeq
where  $a_k$ correspond to the zeros of the Airy function.

\subsubsection{Higher plateau}

On the other hand, the higher plateau should also generate the non-decoupling quasinormal modes. The solution should have $\omega_c>|\hat{\omega}|>\bar{\omega}_c$ such that the wave is purely outgoing with no reflection in the overlapping region, but gets reflected in the far region due to the presence of the second plateau.  Because the second plateau is  in the far-region the usual far-region solution \eqref{equation322} still works, and the non-decoupling modes are determined by the far-region solutions. Although the plateau is beyond the near-region, through a variable replacement $\sR\to\bar{\sR} n^4$ we can pull it back to the near region. For example, after the replacement the leading order  tensor potential becomes
\beq
\frac{\bar{V}_T}{n^2}=\frac{16(1+2\hat{\ell})^2\bar{\sR} ^4+12(5+16\hat{\ell}+16\hat{\ell}^2)\bar{\sR}^2 \alpha^4+24(1+3\hat{\ell}+3\hat{\ell}^2)\bar{\sR} \alpha^6+(1+8\hat{\ell}+8\hat{\ell}^2)\alpha^8+48\bar{\sR}^3 (\alpha+2\hat{\ell}\alpha)^2}{16(\bar{\sR}+\alpha^2)^2(2\bar{\sR}+\alpha^2)^2}.
\eeq
In the limit that $\bar{\sR} \to\infty$, $\bar{V}_T\to n^2\omega_c^2$ as we expected. Near the edge of the second plateau $\bar{\sR}\ll1$, this corresponds $\sR\gg1$ so the solution should be connected with far-region solution of $\Psi_T(\sR)$. We will not illustrate this point in detail since the most important information is the amplitude ratios in front of wave-function components. The far-region solution tells us that in the case of $|\hat{\omega}|>\bar{\omega}_c$ the quasinormal modes should be identical to the ones in the Schwarzschild case, so for the least-damped modes the frequency spectrum is
\beq
\omega_{\text{R}}=\frac{n}{2}+\ell-\frac{a_k}{2^{4/3}}\biggl(\frac{n}{2}+\ell \biggl)^{1/3},
\eeq
and
\beq
\omega_{\text{I}}=-\frac{\sqrt{3}a_k}{2^{4/3}}\biggl(\frac{n}{2}+\ell \biggl)^{1/3}.
\eeq

As a conclusion we find that the interesting things happen when the GB coupling $\tilde{\alpha}$ is very large. There are two kinds of non-decoupling quasinormal modes, one kind is the same as the one in the Schwarzschild black holes. This kind of modes is universal for all asymptotically flat static black holes. The other kind is special for the GB black holes due to the emergence of a new plateau in the potential when $\tilde{\alpha}$ is large enough.

\section{Decoupled modes for gravitational perturbations}\label{section5}

The decoupled modes are normalizable in the near horizon geometry. They are localized within the near horizon region and decoupled with the asymptotically flat region. Their frequencies are of order one. To leading order in $1/n$, these modes are static and becomes dynamical at the next-to-leading order. They can be studied in the $1/n$ expansion order by order.

The  form of the master equation can be recast into the form
\begin{equation}
(L+U_s)\Psi_s(R)=0,
\end{equation}
where
\begin{equation}
L\Psi_s=-\frac{1}{n^2}f\frac{d}{dr}\biggl(f\frac{d}{dr}\Psi_s\biggl),
\end{equation}

\begin{equation}
U_s=\frac{1}{n^2}\biggl(V_s(R)-\omega^2\biggl).
\end{equation}
All quantities can be expanded in powers of $1/n$ as
\begin{equation}
\Psi_s=\sum_{k\ge0}\frac{\Psi^{(k)}_s}{n^k},\qquad L=\sum_{k\ge0}\frac{L^{(k)}}{n^k},\qquad \omega=\sum_{k\ge0}\frac{\omega^{(k)}}{n^k},
\end{equation}
such that the decoupled modes can be studied perturbatively.

The equation for the perturbation at each order is determined by the differential equation with a source
\begin{equation}
\biggl(L^{(0)}+U^{(0)}_s\biggl)\Psi^{(k)}_s=S^{(k)}.
\end{equation}
  Here the sources $\mc{S}^{(k)}$ are obtained from  $L^{(j)}+U_s^{(j)}$ with $j\leq k$, and from the solutions $\Psi_s^{(j)}$ with $j<k$.

At each order, the solution should be normalizable. The strategy to read the decoupled modes is to first compute the lowest order solution and then compute the higher order solution order by order.

\subsection{Small $\tilde{\alpha}$ }
In this case, to the leading order
\begin{equation}
L^{(0)}\Psi=-(\sR-1)\frac{d}{d\sR}\biggl((\sR-1)\frac{d}{d\sR}\Psi \biggl).
\end{equation}
 The equation is exactly the same as the Schwarzschild case, and the Gauss-Bonnet effect is absent. At the next-to-leading order there are corrections from the Gauss-Bonnet term
\beq
L^{(1)}\Psi=2(\sR-1)^2(\frac{\beta}{\sR}+\ln\sR)\Psi''+(\sR-1)(\frac{\beta}{\sR^2}+\frac{\beta-1}{\sR}+1+2\ln \sR)\Psi'.
\eeq

As the decoupled modes are normalizable, the boundary condition
 at $\sR\gg1$ is
\beq
\Psi(\sR\to \infty)\to\frac{1}{\sqrt{\sR}}.
\eeq
This is because the maximum of all the three potentials is $V_s^{\textrm{max}}\to n^2/4$, with $\ell=\mc{ O}(1)$, in the overlapping region, where $1\ll\sR\ll e^n$. The master equation now has the form
\beq
\big(\frac{d^2}{d(\ln \sR)^2}-(\frac{1}{4}-\frac{\omega^2}{n^2})\big)\Psi_s=0.
\eeq
For the decoupled modes $\omega=\mc{O}(1)$, the normalizability of the solution requires $\Psi \sim 1/\sqrt{\sR}$. The other solution being proportional to $\sqrt{\sR}$  is non-normalizable and is excluded.

The boundary condition at the event horizon $\sR=1$ is required by its regularity. This asks the solution to be
\beq
\Psi(\sR \to 1)=e^{-i\omega r_*},
\eeq
where $r_*$ is the tortoise coordinate. Expanding $\Psi(\sR)$ in series of $1/n$, the explicit forms at each order are respectively
\beqa
\Psi^{(0)}(\sR\to 1)&\to&  1\,,\\
\Psi^{(1)}(\sR\to 1)&\to& -i\omega_{(0)}\ln(\sR -1)\,,\\
\Psi^{(2)}(\sR\to 1)&\to& -i(\beta\omega_{(0)}+\omega_{(1)}) \ln(\sR -1)-\frac1 2\omega_{(0)}^2\big((\ln(\sR -1)\big)^2\,,\\
\Psi^{(3)}(\sR\to 1)&\to&i\big(\beta^2\omega_{(0)}-\beta(2\omega_{(0)}+\omega_{(1)})-\omega_{(2)}\big)\ln(\sR-1)-
\omega_{(0)}(\beta\omega_{(0)}+\omega_{(1)})(\ln(\sR-1))^2\nonumber\\
&\,&+\frac{1}{6}i\omega_{(0)}^3\big(\ln(\sR-1)\big)^3,
\eeqa
etc. Note that there are corrections from the GB term in $\Psi^{(i)}, i\geq 2$.

\subsubsection{Tensor type }

For the tensor-type perturbation, the leading order potential is
\begin{equation}
U^{(0)}_T=\frac{\sR^2-1}{4\sR^2},
\end{equation}
which is the same as the Schwarzschild black hole. The solutions are
\begin{equation}
u_0=\sqrt{\sR}, \qquad v_0=\sqrt{\sR}\ln(1-\sR^{-1}).
\end{equation}
Obviously, neither of the two solutions can satisfy the two boundary conditions simultaneously, so there is no decoupled quasinormal mode of the tensor type.

\subsubsection{Vector type }
The vector potential at the leading order is given by
\begin{equation}
U_V^{(0)}=\frac{(\sR-1)(\sR-3)}{4\sR^2},
\end{equation}
The two independent solutions are
\beq
u_0=\frac{1}{\sqrt{\sR}},\qquad v_0=\frac{\sR+\ln(\sR-1)}{\sqrt{\sR}}.
\eeq
The two boundary conditions determine that
\beq
\Psi_V^{(0)}=u_0.
\eeq
At the next-to-leading order, the potential is given by
\beq
U_V^{(1)}=-\frac{(\sR-1)\big(2\beta(\sR-2)+\sR(3-2\ell\sR)+\sR(\sR-3)\ln\sR \big)}{2\sR^3}.
\eeq
The solution that satisfies the boundary condition at the infinity is
\beq
\Psi_V^{(1)}=A_1 u_0-\frac{(\ell-1)\ln(\sR-1)+\ln\sqrt{\sR}}{\sqrt{\sR}},
\eeq
where $A_1$ is an integral constant. The boundary condition at the horizon requires $A_1=0$ and determines the frequency of the decoupled mode to be
\beq
\omega_{(0)}=-i(\ell-1).
\eeq
 Therefore at the next-to-leading order even though the vector potential is modified by the   Gauss-Bonnet term, the quasinormal modes are the same as the ones in the Schwarzschild case.

 At the order in $(1/n)^2$, there are decoupled modes with the frequencies
\beq
\omega_{(1)}=-i(\ell-1)^2+i(\ell-1)\beta.
\eeq
Now, there is a correction from the Gauss-Bonnet term. This conforms to the expansion  form of $f(r)$ when $\tilde{\alpha}$ is small. At the order in $(1/n)^3$, the calculation is straightforward and leads to
\beq
\omega_{(2)}=-i2(\ell-1)^2(\frac{\pi^2}{6}-1)- i \beta(\ell-1)\big(-1+3\beta-3\ell\big).
\eeq

\subsubsection{Scalar type }

Like the situation of the Schwarzschild black hole, in order to properly deal with the region where $\sR=\mc{O}(n)$ we need to introduce a new variable
\beq
\bar{\sR}=\frac{\sR}{n}.
\eeq
Then the potential to the leading order becomes
\beq
V_S(\bar{\sR})=n^2\bar{V}_S(\bar{\sR})=\frac{n^2}{4}\frac{1-12(\ell-1)\bar{\sR}+4(\ell-1)^2\bar{\sR}^2}{(1+2(\ell-1)\bar{\sR})^2}.
\eeq
 This potential correctly captures all the features of the scalar potential in the near-region. Especially, in the region $1\ll\sR\ll n$, we have a small $\bar{\sR}$ which should be matched with the solution of $\Psi(\sR)$ for $\sR=\mc O(1)$.

First of all it is straightforward to find the solutions for $\Psi_S(\sR)$ with the ingoing boundary condition at the horizon
\beq
\Psi_S^{(0)}(\sR)=\sqrt{\sR},
\eeq
\beq
\Psi_S^{(1)}(\sR)=\sqrt{\sR}\big(-2(\ell-1)(\sR-1)-i\omega_{(0)}\ln(\sR-1)+(1-2\ell+2i\omega_{(0)})\ln\sqrt{\sR}\big),
\eeq
Up to the second order, at the large $\sR$ the expansion of $\Psi_S(\sR)$ gives
\beq\label{equation429}
\Psi_S(\sR)=\sqrt{\sR}\big[1+\frac{1}{n}\big(\frac{i\omega_{(0)}}{\sR}+2(\ell-1)-2(\ell-1)\sR -(2\ell-1)\ln\sqrt{\sR}\big) \big].
\eeq

To the second order, the solution for $\bar{\Psi}_S(\bar{\sR})$ with the boundary condition $\bar{\Psi}_S(\bar{\sR})\sim 1/\sqrt{\bar{\sR}}$ as $\bar{\sR}\to \infty$ is
\beqa
\bar{\Psi}_S^{(0)}(\bar{\sR})+\frac{1}{n}\bar{\Psi}_S^{(1)}(\bar{\sR})=\frac{C_1\sqrt{\bar{\sR}}}{1+2(\ell-1)\bar{\sR}}[1-\frac{1}{n}\big(\frac{3+2\beta+(2-6\ell+4\ell^2)\bar{\sR}}{2+4(\ell-1)\bar{\sR}}-(2\ell-1)\ln\sqrt{\bar{\sR}} \big)].\nonumber
\eeqa
The match of two solutions at the leading order requires $C_1=\sqrt{n}+B_1/\sqrt{n}$ with $B_1$ being undetermined, but this is not sufficient to determine the frequency of the decoupled mode because there is  $1/\bar{\sR}$ term coming from $\bar{\Psi}_S^{(2)}(\bar{\sR})/n^2$. Indeed there is such a term
\beq
\frac{1}{n}\frac{\ell^2-\ell+\omega_{(0)}^2}{2(\ell-1)\sqrt{\sR}},
\eeq
which, by matching with \eqref{equation429}, determines that
\beq
\omega_{(0)\pm}=\pm\sqrt{(\ell-1)}-i(\ell-1).
\eeq
This is equal to the result of the Schwarzschild black hole. One can proceed to find  the frequency of the decoupled mode  in the $1/n$ order
\beq
\omega_{(1)\pm}=\pm\sqrt{(\ell-1)}(\frac{3\ell}{2}-2-\beta)-i(\ell-1)(\ell-2-\beta),
\eeq
which encodes the correction from the Gauss-Bonnet term. The discussion for the higher order decoupled modes is similar and straightforward but becomes more and more complicated.

%Although the proceeding calculation is identical but the series expansion of scalar potential is too complicated for us to continue, so we stop here.
\subsection{Large $\tilde{\alpha}$ }
In this case, there is
\begin{equation}
L^{(0)}\Psi=-(1-\sR^{-\frac{1}{2}})\sR\frac{d}{d\sR}\biggl((1-\sR^{-\frac{1}{2}})\sR\frac{d}{d\sR}\Psi\biggl).
\end{equation}
As in the situation of small $\tilde{\alpha}$, the boundary condition at the horizon can be  given order by order as
\beqa
\Psi^{(0)}(\sR\to 1)&\to& 1,\\
\Psi^{(1)}(\sR\to 1)&\to& -2i\omega_{(0)} \ln(\sqrt{\sR }-1),\\
\Psi^{(2)}(\sR\to 1)&\to& -2i(\omega_{(1)}+ 2\omega_{(0)})\ln(\sqrt{\sR}-1)-2\omega_{(0)}^2(\ln(\sqrt{\sR }-1))^2,\\
\Psi^{(3)}(\sR\to 1)&\to&-2i(\omega_{(2)}+2\omega_{(1)}+(4-\frac{1}{\alpha})\omega_{(0)})\ln(\sqrt{\sR}-1)
\nonumber\\
&\,&-4\omega_{(0)}(2\omega_{(0)}+\omega_{(1)})(\ln(\sqrt{\sR}-1))^2+\frac{4}{3}i\omega_{(0)}^3\lp\ln(\sqrt{\sR }-1)\rp^3,
\eeqa
etc. The boundary condition at $\sR\gg1$ requires
\beq\label{euqa}
\Psi(\sR\to\infty)\to \frac{1}{\sR^{1/4}}.
\eeq
Note that this is different from the Schwarzschild case,  because the maximum of all the three potentials in the large $\tilde \alpha$ case is $V_s^{\textrm{max}}\to n^2/16$. In order to be normalizable,
the decoupled modes  should satisfy \eqref{euqa}.

\subsubsection{Tensor type }
The leading order of the tensor potential $U_T$ is
\beq
U_T^{(0)}=\frac{\sR-1}{16\sR},
\eeq
and the corresponding solutions are
\beq
u_0=\sR^{1/4}, \qquad v_0=\sR^{1/4}\,\ln(1-\frac{1}{\sqrt{\sR}}).
\eeq
Obviously, none of the two solutions can satisfy the two boundary conditions simultaneously, and there is no decoupled quasinormal modes of tensor type.

\subsubsection{Vector type }

\begin{figure}[t]
 \begin{center}
  \includegraphics[width=.5\textwidth,angle=0]{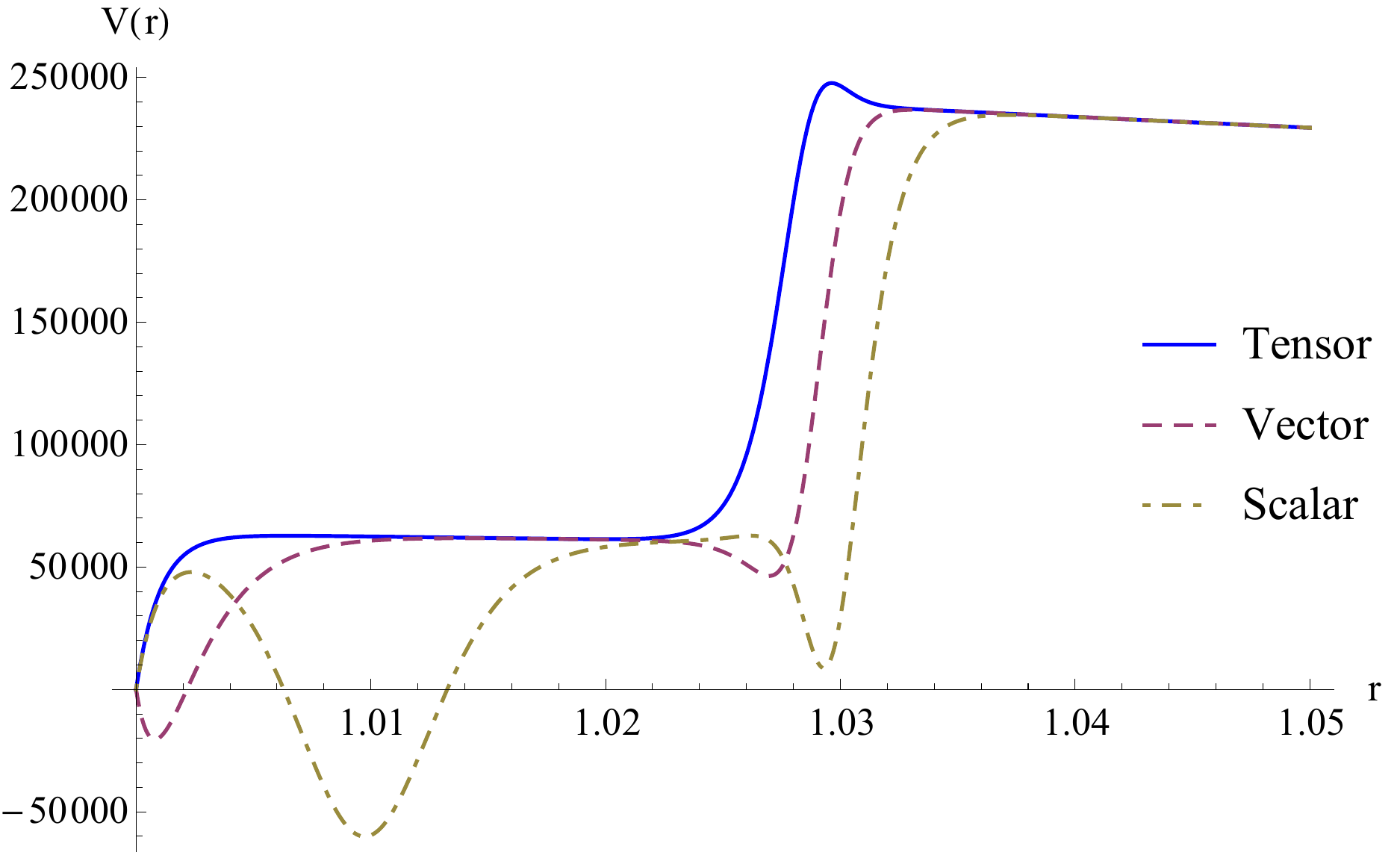}
   \end{center}
 \vspace{-5mm}
 \caption{\small Radial potential $V_s(r)$ for the perturbations of the Einstein-Gauss-Bonnet black hole for $n=10^3$, $\ell=3$ and $\tilde{\alpha}=10^6$ . The horizon is at $r_H=1$. Note that there are two minima for the vector and scalar potential.  }
 \label{2}
\end{figure}
The leading order of the vector potential $U_V$ is given by
\beq\label{equation542}
U_V^{(0)}=\frac{3-4\sqrt{\sR}+\sR}{16\sR},
\eeq
and the two independent solutions are
\beq
u_0=\frac{1}{\sR^{1/4}}, \qquad v_0=\frac{\sqrt{\sR}+\ln(\sqrt{\sR}-1)}{\sR^{1/4}}.
\eeq
The boundary conditions select
\beq
\Psi_V^{(0)}=u_0,
\eeq
so there could exist the quasinormal modes of vector type. At the next-to-leading order,
\beq
U_V^{(1)}=\frac{2(-1+2\ell)(-1+\sqrt{\sR})-(-2+\sqrt{\sR})\ln\sR}{8\sqrt{\sR}},
\eeq
then  the solution is
\beq
\Psi_V^{(1)}=C_1\,u_0+C_2\,\frac{\sqrt{\sR}+\ln(\sqrt{\sR}-1)}{\sR^{1/4}}-\frac{2(\ell-1)\ln(\sqrt{\sR}-1)}{\sR^{1/4}},
\eeq
where $C_1$ and $C_2$ are two integration constants. The boundary condition at the infinity  requires $C_1=C_2=0$, and the boundary condition at the horizon determines the frequency to be
\beq
\omega_{(0)}=-i(\ell-1).
\eeq
It is a surprise that this is exactly the same as the one in the Schwarzschild case found in \cite{kw8}.  At the second order in $1/n$, the potential is
\beq
\begin{split}
U_V^{(2)}=\frac{1}{8\alpha\sR}\biggl[&(\sqrt{\sR}-1)\big(-3+\sqrt{\sR}(1+4(\ell-1)^2\alpha)\big)-2(2\ell-1)(2\sR-\sqrt{\sR})\alpha\ln\sR
\\&+(\sR-\sqrt{\sR})\alpha(\ln\sR)^2\biggl]-\omega_{(0)}^2.
\end{split}
\eeq
Then the solution can be easily obtained.  With the help of the boundary condition at the  horizon,  the second order frequency can be read
\beq
\omega_{(1)}=-i(\ell-1)(\ell-3),
\eeq
which is different with the one in the Schwarzschild case due to different effect from the boundary conditions. Actually, there is $\omega_{(1)}+ 2\omega_{(0)}$ instead of a simple $\omega_{(1)}$ from the ingoing boundary condition, and the quasinormal mode at the second order is accordingly changed. At the third order, it is straightforward to read
\beq
\omega_{(2)}=-i4(\ell-1)^2(\frac{\pi^2}{3}-1).
\eeq

From Fig.~\ref{2} we see that the vector potential has two minima, the negative one on the left side and the positive one on the right side. The negative one gives the above decoupled mode. The positive one is  new, which  does not appear when
$\tilde{\alpha}$ is small, and its location is at $\sR\sim\mc O(n^4) $. We would like to investigate whether the positive one gives new decoupled modes or it leads to the same wave function discussed above propagating through the whole regions. In order to properly deal with the region with positive minimum, we introduce a new variable

\beq
\bar{\sR}=\frac{\sR}{n^4}.
\eeq
Since $\bar{\sR}\sim\mc O(1)$ in the positive minimum region, we can study the wave function around the minimum using the $1/n$ expansion.
The leading order vector potential in terms of $\bar{\sR}$ is
\beq
\bar{U}_V^{(0)}(\bar{\sR})=\frac{16\bar{\sR}^4+48\bar{\sR}^3\alpha^2+28\bar{\sR}^2\alpha^4+\alpha^8}{16(\bar{\sR}+\alpha^2)^2(2\bar{\sR}+\alpha^2)^2}.
\eeq
In the limit that $\bar{\sR}$ is very small the potential reaches $1/16$ which can be matched to the maximum of the  potential \eqref{equation542} in the region $1\ll\sR\ll \tilde{\alpha}^2$. And when $\bar{\sR}$ is very large
the potential has a maximum $1/4$ which is the same as the Schwarzschild black hole,  since the spacetime is asymptotically flat.

At the leading order the differential equation becomes
\beq
\bar{\sR}^2\bar{\Psi}_V''(\bar{\sR})+\bar{\sR}\bar{\Psi}_V'(\bar{\sR})-\bar{U}_V^{(0)}(\bar{\sR})\bar{\Psi}_V(\bar{\sR})=0,
\eeq
and its solutions are
\beq
\bar{u}_0=\frac{(\bar{\sR}+\alpha^2)^{1/4}}{\bar{\sR}^{1/4}\sqrt{2\bar{\sR}+\alpha^2}},
\qquad \bar{v}_0=\frac{2\bar{\sR}^{1/4}(\bar{\sR}+\alpha^2)^{3/4}}{\sqrt{2\bar{\sR}+\alpha^2}}.
\eeq
At the large $\bar{\sR}$, $\bar{u}_0$ gives the correct asymptotic behavior which is $1/\sqrt{\bar{\sR}}$. At the small $\bar{\sR}$, $\bar{u}_0$ scales as $1/\sR^{1/4}$, this can be matched to the solution $\Psi_V^{(0)}$ in the region
 $1\ll\sR\ll \tilde{\alpha}^2$.

 Next let us extend the
  discussion to the next-to-leading order. The solution $\Psi_V$ with the ingoing boundary condition at the horizon is
\beqa
\Psi_V(\sR)&=&\Psi_V^{(0)}(\sR)+\frac{1}{n}\Psi_V^{(1)}(\sR)\nonumber\\
&=&\frac{1}{\sR^{1/4}}\big[1+\frac{1}{n}\big(2i\omega-2(\ell-1)+2(\ell-1-i\omega)\sR^{1/2}-2i\omega\ln(\sqrt{\sR}-1)\big)\big].
\eeqa
Its large $\sR$ behavior is
\beq\label{equation556}
\Psi_V(\sR)=\frac{1}{\sR^{1/4}}\big[1+\frac{1}{n}\big(\frac{2i\omega}{\sqrt{\sR}}+2i\omega-2(\ell-1)+2(\ell-1-i\omega)\sR^{1/2}-2i\omega\ln(\sqrt{\sR})\big)\big].
\eeq
On the other hand, to the same order the solution for $\bar{\Psi}_V(\bar{\sR})$ with the correct boundary condition at the large $\bar{\sR}$ is
\beqa
&\,&\bar{\Psi}_V^{(0)}(\bar{\sR})+\frac{1}{n}\bar{\Psi}_V^{(1)}(\bar{\sR})=\frac{C_1}{2\bar{\sR}^{1/4}(\bar{\sR}+\alpha^2)^{3/4}(2\bar{\sR}+\alpha^2)^{3/2}}\biggl[
2(\bar{\sR}+\alpha^2)(2\bar{\sR}+\alpha^2)\nonumber\\
&&+\frac{1}{n}\biggl(\frac{41(\ell-1)\sqrt{\bar{\sR}}(\bar{\sR}+\alpha^2)^{3/2}(2\bar{\sR}+\alpha^2)}{\alpha^2}\nonumber \\
&&-\frac{1}{\alpha^2}(-8\bar{\sR}^3+8\ell \bar{\sR}^3-18\bar{\sR}^2\alpha^2+20\ell \bar{\sR}^2\alpha^2-15\bar{\sR}\alpha^4+16\ell \bar{\sR}\alpha^4-6\alpha^6+4\ell \alpha^6)\nonumber\\
&&- \big((4\ell-2)\bar{\sR}^2+3(2\ell-1)\bar{\sR}\alpha^2+2(\ell-1)\alpha^4 \big)\ln\bar{\sR}+4\alpha^4\ln n\biggl)\biggl],
\eeqa
where $C_1$ is an integral constant.  Then we make a replacement $\bar{\sR}\to \sR/n^4$ and expand the expression in $1/n$. From the matching with \eqref{equation556} at the leading order, we can fix $C_1=\sqrt{\alpha}/n+B_1/n^2$. To the next-to-leading order we get
\beq
\bar{\Psi}_V^{(0)}(\bar{\sR})+\frac{1}{n}\bar{\Psi}_V^{(1)}(\bar{\sR})=
\frac{1}{\sR^{1/4}}\biggl[1+\frac{1}{n}(\frac{B_1}{\sqrt{\alpha}}+3-2\ell-(\ell-1)\ln\sR+2(2\ell-1)\ln n) \biggl].
\eeq
The matching with \eqref{equation556} can determine all the undetermined constants ($1/\sqrt{\sR}$ term comes from the next order), among which we have
\beqa
\omega=-i(\ell-1).
\eeqa
Hence this verifies that the positive minimum do not give any new decoupled mode. Actually, the wave in the left valley of the potential propagate right to the next valley. In other words, once we find the solution in the left valley, we can extend it to the right and using the matching condition in the overlapping region we can determine the wavefunction in the right valley completely.

It seems that the decoupled modes are only determined by the wavefunction in the left potential valley with the asyptotically boundary condition $\Psi(\sR \to \infty) \sim (1/\sR)^{1/4}$. This is due to the fact that the potential plateau between two minima has a long enough extension. As we discussed above, the wavefunction in the second valley could be determined by the matching of the solution.
A better treatment is to find the solutions in different regions and paste them correctly, and then read the frequency of the decoupled modes. We will have to use this treatment for the scalar type perturbation in the next subsection.

\subsubsection{Scalar type }

We follow the similar treatment in the small $\tilde{\alpha}$ case. However now the first  minimum locates at $\sR=\mc O(n^2)$, so the correct variable  should be
\beq
\bar{\sR}=\frac{\sR}{n^2},
\eeq
then the scalar potential in the leading order becomes
\beq
\bar{V}_S(\bar{\sR})=\frac{n^2}{16}\frac{1-24(\ell-1)\sqrt{\bar{\sR}}+16(\ell-1)^2\bar{\sR}}{(1+4(\ell-1)\sqrt{\bar{\sR}})^2}.
\eeq
From this expression we can read all the features appearing in  Fig.~\ref{2}: it reaches the same maxima $n^2/16$ at the small $\bar{\sR}$ and the large $\bar{\sR}$, and reaches a minimum between these two maxima at $\bar{\sR}=1/16(\ell-1)^2$. Note that the scalar potential has a local maximum before the minimum.

To determine the decoupled modes, we need to find the solutions in different regions and paste them correctly. First let us first  match the solutions for  $\bar{\Psi}_S(\bar{\sR})$ and the solutions for $\Psi_S(\sR)$ in the region $1\ll\bar{\sR}\ll n^2$. It is easy to find the solutions for $\Psi_S(\sR)$. With the ingoing boundary condition at the horizon, we get
 \beq
 \Psi_S^{(0)}(\sR)=\sR^{1/4},
 \eeq
 \beq
  \Psi_S^{(1)}(\sR)=\sR^{1/4}\big[ -4(\ell-1)(\sqrt{\sR}-1)-2i\omega_{(0)}\ln(\sqrt{\sR}-1)-(\ell-i\omega_{(0)})\ln\sR\big].
 \eeq
 In the large $\sR$ region,  $\Psi_S$ becomes
 \beqa\label{equation464}
\Psi_S(\sR)&=&\Psi_S^{(0)}(\sR)+\frac{1}{n}\Psi_S^{(1)}(\sR)\nonumber\\
&=&\sR^{1/4}\big[1+\frac{1}{n}\big(\frac{i2\omega_{(0)}}{\sqrt{\sR}} -4(\ell-1)(\sqrt{\sR}-1)-\ell\ln\sR\big)\big].
\eeqa

On the other hand, up to the next-to-leading order the solution for $\bar{\Psi}_S(\bar{\sR})$ with the boundary condition $\bar{\Psi}_S(\bar{\sR})\sim \bar{\sR}^{-1/4}$  is
\beqa
\bar{\Psi}_S^{(0)}(\bar{\sR})+\frac{1}{n}\bar{\Psi}_S^{(1)}(\bar{\sR})=\frac{C_0\bar{\sR}^{1/4}}{1+4(\ell-1)\sqrt{\bar{\sR}}}
\left(1-\frac{1}{n}\big(2(\ell-1)(1+\frac1 2\ln\bar{\sR})+\frac{2-\ell+\ln\bar{\sR}+2\ln n}{1+4(\ell-1)\sqrt{\bar{\sR}}} \big)\right),\nonumber\\
\eeqa
Comparing with \eqref{equation464}, we find that the matching at the leading order requires that $C_0=\sqrt{n}+C_1/\sqrt{n}$ as before and the $\omega_{(0)}$ term is given at the next order in $\bar{\Psi}_S^{(2)}(\bar{\sR})/n^2$. However, a little subtlety here is that the third order wavefunction $\bar{\Psi}_S^{(2)}(\bar{\sR})$ is not convergent any more for large $\bar{\sR}$, and it seems that the far region boundary condition cannot be achieved. However, there is nothing bad that truly happens. This can be explained by examining Fig.~\ref{2} carefully. The figure shows clearly that the scalar potential is very different from the vector and the tensor potentials with a remarkable feature that the plateau connecting the two valleys may not be sufficient long and high for the wavefunction to decay into zero. Therefore the true far-region is in the higher plateau which has the same structure as the Schwarzschild black holes. For the first and the second order wave functions the lower plateau is long enough so that they would not stretch into the higher plateau, but for the third and higher order wave functions the wave stretches into farther region. Therefore we need to discuss the wavefunction in the second valley carefully.

To investigate the wave in the second valley, we introduce the variable
\beq
\tilde{\sR}=\frac{\sR}{n^4}.
\eeq
  It can be used to investigate the region $\sR\sim \mc O(n^4)$ which is  the location of the second valley and the edge of the higher plateau. The leading order potential is now
\beq
\tilde{V}_S(\tilde{\sR})=\frac{n^2}{16}\frac{16\tilde{\sR}^4-16\tilde{\sR}^3\alpha^2-4\tilde{\sR}\alpha^4+8\tilde{\sR}\alpha^6+\alpha^8}{(\tilde{\sR}+\alpha^2)^2(2\tilde{\sR}+\alpha^2)^2}.
\eeq
In the limit $\tilde{\sR}\to 0$, $\tilde{V}_S\to n^2/16$ which is the height of the middle plateau. Moreover when  $\tilde{\sR}\to \infty$, $\tilde{V}_S\to n^2/4$  gives the correct  far-region behavior of GB black holes. The  wave functions  $\tilde{\sR}$ should satisfy the boundary condition
\beq\label{equation569}
\tilde{\Psi}_S(\tilde{\sR}\to \infty)\to \frac{1}{\sqrt{\tilde{\sR}}}.
\eeq
With the potential and the boundary condition, the leading order solution is
\beq
\tilde{\Psi}_S^{(0)}(\tilde{\sR})=\frac{\sqrt{2\tilde{\sR}+\alpha^2}}{\tilde{\sR}^{1/4}(\tilde{\sR}+\alpha^2)^{3/4}}.
\eeq
 On the other hand, at a small $\tilde{\sR}=\bar{\sR}/n^2$ we find $\tilde{\Psi}_S^{(0)}(\tilde{\sR})\to1/\bar{\sR}^{1/4}$ so that it can be matched to the solution $\bar{\Psi}_S^{(0)}(\bar{\sR})$. %Thus, even if the wavefunction cannot be zero when $\bar{\sR} \to \infty$, it does not have to be since that is not the true infinity and in the true infinity, the boundary condition is indeed satisfied.

Therefore we have three pieces of wave functions: the first is $\Psi(\sR)$ which satisfies the ingoing boundary condition at the event horizon, the second is $\bar{\Psi}_S(\sR)$ which is valid in the first valley in Fig.~\ref{2} and matches with $\Psi(\sR)$ in the region $1\ll \sR\ll n^2$, the third is  $\tilde{\Psi}_s(\tilde{\sR})$ which is valid in the second valley and satisfies the far-region boundary condition \eqref{equation569} and can be matched with $\bar{\Psi}_S(\sR)$ in the region $1\ll \bar{\sR}\ll n^2$. The three pieces should be matched in the overlapping region, which constrains all of the undetermined constants in solving the differential equation. This computation can be carried out order by order. At the first order all the boundary conditions can be satisfied provided that we have
\beq
\omega_{(0)}=\pm\sqrt{\ell-1}-i(\ell-1).
\eeq
At the next order, the frequency is
\beq
\omega_{(1)}=\pm\sqrt{\ell-1}(\frac{3\ell}{2}-4)-i(\ell-1)(\ell-4).
\eeq
Similar to the vector-type decoupled modes, the scalar decoupled mode at the first order is the same as the one in the Schwarzschild case, but the scalar mode at  the second order is different. %The discrepancy in the second order is entirely due to the difference in the boundary condition.

\section{Summary and discussions}

In this paper we  studied the  quasinormal modes of the Gauss-Bonnet black holes in the large D.  We have obtained the quasinormal spectrum of a minimally coupled scalar field in the background of the Gauss-Bonnet black hole and  three types of quasinormal modes of gravitational perturbations. Since the metric expansion depends on the value of the GB parameter $\tilde{\alpha}$, we chose two typical values, a small $\tilde{\alpha}$ of order $1/n$ and a large $\tilde{\alpha}$ of order $n^2$ to investigate. In the large D limit, the geometry of the Gauss-Bonnet black hole is qualitatively similar to the one of the Schwarzschild case: the near horizon region becomes very short and approach flat spacetime very quickly.

For the scalar field the quasinormal modes are identical to the ones in the Schwarzschild case\cite{kw8}, and they are independent of the coupling constant $\tilde{\alpha}$. This is in accord with the fact that the scalar quasinormal modes are universal and is insensitive to the black hole geometry.

When the effective GB parameter $\tilde{\alpha}$ is small,  the non-decoupling modes of the gravitational perturbations are identical to the ones in the Einstein gravity. This is
easy to understand as the effect of the GB term is negligible. However, when the effective GB parameter $\tilde{\alpha}$ is large,  there is another set of decoupling quasinormal modes, besides the ones in the Schwarzschild case. This is due to the appearance of another plateau in the radial potential. The basic picture is that even if the GB black hole and Schwarzschild black holes share the same asymptotic geometry, the near region geometry is slightly different  so that the non-decoupling modes in two cases are slightly different.
Nevertheless, all the non-decoupling modes are non-renormalizable in the near horizon geometry, and their frequencies $\omega \sim D$.

For the decoupled modes, when the parameter $\tilde{\alpha}$ is small, the effect of the Gauss-Bonnet term only appears beyond the leading order. This is within  our expectation since the Gauss-Bonnet term is just a small modification to the Einstein's gravity after all. When the parameter $\tilde{\alpha}$ is large, the radial potentials for the vector-type and scalar type present new features: there are two minima rather than one, and the shapes of the potentials for the vector and the scalar are different. There are a few remarkable points:
\begin{enumerate}
  \item There is no tensor-type decoupled mode. This can be seen easily from the potential: there is no place to define a normalizable mode.
  \item For the vector-type perturbation, one can read the decoupled modes from the wavefunction in the first valley, as the plateau between two valleys are long enough.
  \item For the scalar-type perturbation, one has to compute the wavefunction in three regions and paste them correctly to read the decoupled modes.
  \item  At the leading order the frequencies of the decoupled modes are the same as the ones in the small $\tilde \alpha$. This is a little surprise since in this case the Gauss-Bonnet term should be dominant, and it raises an issue if the leading decoupled mode is universal or not. However the numerical analysis of the vector-type modes in the appendix shows that this is just a coincidence and even at the leading order the frequencies are not universal when $\tilde{\alpha}$ takes an intermediate value.
\end{enumerate}

It would be interesting to compare our analytic results with the numerical study. In \cite{kw19}, the quasinormal modes of the GB black holes have been studied numerically in $D=5-11$ for a large $\alpha$. It was found that the instability in $D=5,6$ disappears in larger $D$. To compare with our results obtained in this paper, one has to push the study to much larger $D$.

The study of the quasi-normal modes in the Gauss-Bonnet black hole at a large D is the first step to understand the black hole dynamics in the Gauss-Bonnet gravity. The decoupled modes encodes the nontrivial black hole physics. It would be interesting to extend the study to the nonlinear regime, as suggested recently in \cite{kw10,Bhattacharyya:2015fdk}.

\section*{Acknowledgments}

The work was in part supported by NSFC Grants No.~11275010, No.~11335012 and No.~11325522.
 BC would like to thank the participants of the advanced workshop ``Dark Energy and Fundamental Theory" supported by the Special Fund for Theoretical Physics from the National Natural Science Foundations of China with Grant No. 11447613 for stimulating discussion.

\appendix
\section{Numerical results of the first leading order of decoupled quasinormal modes of ``hybrid" Gauss-Bonnet black holes}

The computation of the decoupled modes in the Gauss-Bonnet black holes when $\tilde{\alpha}$ is of order one  is similar , but the analytical results are difficult to obtain. Therefore, one has to use numerical method to find the solution of ordinary differential equations with suitable boundary conditions. The large $D$ expansion can still simplify the numerical calculation dramatically. Here we list  the decoupled quasinormal modes of the vector-type perturbations at the leading order in Table \ref{1}. 

\begin{table}[h]
  \centering
  
    \begin{tabular}{|c|c|c|c|}
\hline
$\omega(\tilde{\alpha}=0.05)$&$\omega(\tilde{\alpha}=0.1)$&$\omega(\tilde{\alpha}=0.15)$\\\hline
$-\mathrm{i}(0.956371l-0.956001)$&$-\mathrm{i}(0.924064l-0.923859)$&$-\mathrm{i}(0.899545l-0.899402)$\\\hline
$\omega(\tilde{\alpha}=0.2)$&$\omega(\tilde{\alpha}=0.25)$&$\omega(\tilde{\alpha}=0.3)$\\\hline
$-\mathrm{i}(0.880862l-0.880752)$&$-\mathrm{i}(0.866594l-0.866505)$&$-\mathrm{i}(0.855708-0.855632)$\\\hline
$\omega(\tilde{\alpha}=0.35)$&$\omega(\tilde{\alpha}=0.4)$&$\omega(\tilde{\alpha}=0.45)$\\\hline
$-\mathrm{i}(0.847443l-0.847377)$&$-\mathrm{i}(0.841224l-0.841166)$&$-\mathrm{i}(0.836619l-0.836568)$\\\hline

$\omega(\tilde{\alpha}=0.5)$&$\omega(\tilde{\alpha}=0.55)$&$\omega(\tilde{\alpha}=0.65)$\\\hline
$-\mathrm{i}(0.8333l-0.833266)$&$-\mathrm{i}(0.830999l-0.830968)$&$-\mathrm{i}(0.828697l-0.828671)$\\\hline

$\omega(\tilde{\alpha}=0.8)$&$\omega(\tilde{\alpha}=1)$&$\omega(\tilde{\alpha}=1.5)$\\\hline
$-\mathrm{i}(0.82904l-0.829018)$&$-\mathrm{i}(0.833315l-0.833299)$&$-\mathrm{i}(0.849988l-0.849974)$\\\hline
$\omega(\tilde{\alpha}=2)$&$\omega(\tilde{\alpha}=2.5)$&$\omega(\tilde{\alpha}=3)$\\\hline
$-\mathrm{i}(0.866657l-0.866646)$&$-\mathrm{i}(0.880945l-0.880936)$&$-\mathrm{i}(0.892852l-0.892843)$\\\hline
$\omega(\tilde{\alpha}=3.5)$&$\omega(\tilde{\alpha}=4)$&$\omega(\tilde{\alpha}=4.5)$\\\hline
$-\mathrm{i}(0.902774l-0.902765)$&$-\mathrm{i}(0.911107l-0.911099)$&$-\mathrm{i}(0.918178l-0.918172)$\\\hline
    \end{tabular}%
\caption{Decoupled quasinormal Modes for the ``hybrid" Gauss-Bonnet black holes perturbation of the vector type. The value of $\alpha$ and the frequencies are measured in units
of the horizon radius ($r_0=1$).}
  \label{1}%
\end{table}%

\begin{figure}[h]
\centering
\includegraphics[width=7cm]{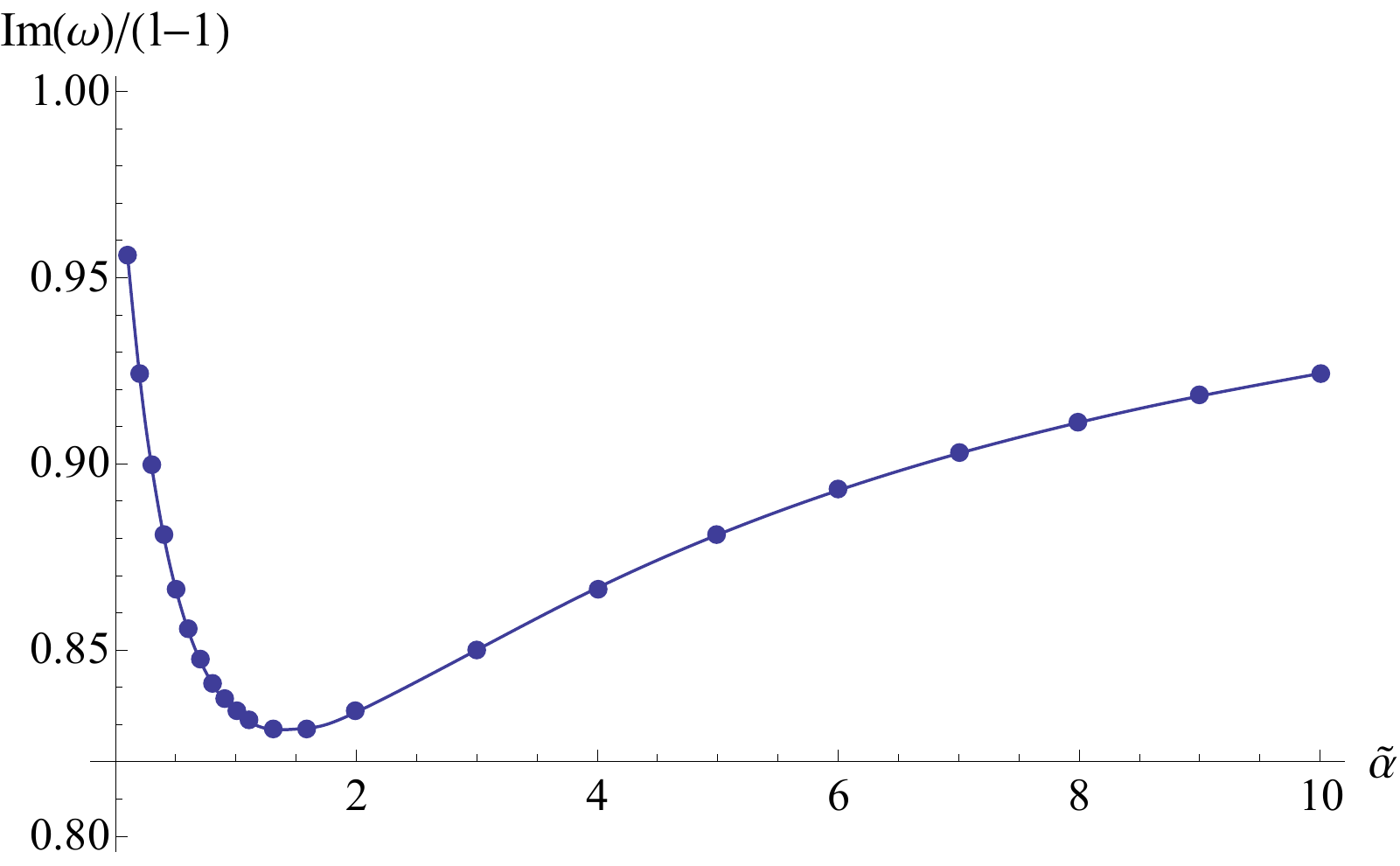}
\caption{The decoupled quasinormal modes at the leading order versus the parameter $\tilde{\alpha}$ for the vector-type perturbation in a ``hybrid" Gauss-Bonnet black hole.}%  Units are $r_0=1$.}
\label{3}
\end{figure}

It is obvious that these decoupled quasinormal modes are all of the form $-\mathrm{i}k(l-1)$, with $k$ varying between 0.8 to 1. The relation between $k$ and $\tilde{\alpha}$ are plotted in Fig. \ref{3}, with a minimum at around $\tilde{\alpha}=0.65$. It indicates that black holes are more ``stable" when the GB term or the Einstein term dominates. On the other hand, when the  GB term and the Einstein term are comparable, the black holes are less ``stable".


\begin{thebibliography}{}

\bibitem{kw3} R. Emparan, R. Suzuki, and K. Tanabe,{\em  The large $D$ limit of General Relativity, JHEP} \textbf{1306} (2013) 009, [arXiv:1302.6382[hep-th]].
\bibitem{kw4} R. Emparan, D. Grumiller, and K. Tanabe, {\em Large-$D$ gravity and low-$D$ strings, Phys.Rev.Lett.} \textbf{110} (2013), no. 25 251102, [arXiv:1303.1995[hep-th]].
\bibitem{kw5} R. Emparan and K. Tanabe, {\em Holographic superconductivity in the large $D$ expansion, JHEP} \textbf{1401} (2014) 145, [arXiv:1312.1108[hep-th]].
\bibitem{kw1} G. 't Hooft, {\em A Planar Diagram Theory for Strong Interactions, Nucl.Phys.} \textbf{B72} (1974) 461.
\bibitem{kw2} E. Witten, {\em Quarks, atoms, and the $1/N$ expansion, Physics Today} \textbf{33} (1980).
    \bibitem{kw20} F. R. Tangherlini, {\em Schwarzschild field in $n$ dimensions and the dimensionality of
space problem, Nuovo Cim.} \textbf{27} (1963) 636.
\bibitem{kw6} R. Emparan and K. Tanabe, {\em Universal quasinormal modes of large $D$ black holes, Phys.Rev.} \textbf{D89} (2014), no. 6 064028, [arXiv:1401.1957[hep-th]].
\bibitem{kw7} R. Emparan, R. Suzuki, and K. Tanabe, {\em Instability of rotating black holes: large $D$ analysis, JHEP} \textbf{1406} (2014) 106, [arXiv:1402.6215[hep-th]].
\bibitem{kw8} R. Emparan, R. Suzuki, and K. Tanabe, {\em Decoupling and non-decoupling dynamics of large $D$ black holes, JHEP} \textbf{1407} (2014) 113, [arXiv:1406.1258[hep-th]].
\bibitem{kw9} R. Emparan, R. Suzuki, and K. Tanabe, {\em Quasinormal modes of (Anti-)de Sitter black holes in the $1/D$ expansion, JHEP} \textbf{04} (2015) 085, [arXiv:1502.02820[hep-th]].
\bibitem{kw10} S. Bhattacharyya, A. De, S. Minwalla, R. Mohanc, and A. Saha, {\em A membrane paradigm at large D}, arXiv:1504.06613 [hep-th].
\bibitem{Giribet:2013wia}
  G.~Giribet,
  ``Large D limit of dimensionally continued gravity,''
  Phys.\ Rev.\ D {\bf 87}, no. 10, 107504 (2013)
  doi:10.1103/PhysRevD.87.107504
  [arXiv:1303.1982 [gr-qc]].
  
\bibitem{Zwiebach:1985uq}
  B.~Zwiebach,
  {\em Curvature Squared Terms and String Theories,}
  Phys.\ Lett.\ B {\bf 156}, 315 (1985).

\bibitem{kw11} D. G. Boulware and S. Deser, {\em String-Generated Gravity Models, Phys. Rev. Lett.} \textbf{55} (1985), no. 24 2656.
%\bibitem{kw12} R. G. Cai, {\em Gauss-Bonnet Black Holes in AdS Spaces, Phys.Rev.}, \textbf{D65} (2002), no. 8 084014, [arXiv:hep-th/0109133].




\bibitem{kw17} G. Dotti and R. J. Gleiser, {\em Linear stability of Einstein-Gauss-Bonnet static spacetimes
Part I: tensor perturbations, Phys. Rev.} \textbf{D72} (2005), no. 4, 044018, [arXiv:gr-qc/0503117].
\bibitem{kw18} R. J. Gleiser and G. Dotti, {\em Linear stability of Einstein-Gauss-Bonnet static spacetimes- Part
II: Vector and scalar perturbations, Phys. Rev.} \textbf{D72} (2005), no. 12, 124002, [arXiv:gr-qc/0510069].
\bibitem{kw15} R. G. Daghigh, G. Kunstatter and J. Ziprick, {\em The Mystery of the Asymptotic Quasinormal Modes of Gauss-Bonnet Black Holes, Class. Quant. Grav.} \textbf{24}, 1981 (2007), [arXiv:gr-qc/0611139].
\bibitem{kw19} R. A. Konoplya and A. Zhidenko, {\em (In)stability of $D$-dimensional black holes in Gauss-Bonnet theory, Phys. Rev.} \textbf{D77} (2008), no. 10, 104004, [arXiv:0802.0267].
\bibitem{Bhattacharyya:2015fdk}
  S.~Bhattacharyya, M.~Mandlik, S.~Minwalla and S.~Thakur,
  {\em A Charged Membrane Paradigm at Large D},
  arXiv:1511.03432 [hep-th].

\end{thebibliography}
\end{document}